\documentclass[twocolumn]{aastex62}
\usepackage{amsmath}
\usepackage{graphicx}

\newcommand{\beq}{\begin{equation}}
\newcommand{\eeq}{\end{equation}}
\newcommand{\bea}{\begin{eqnarray}}
\newcommand{\eea}{\end{eqnarray}}

\begin{document}

\title{\bf Constraints on the binary black hole hypothesis for system LB-1}

\author{Rong-Feng Shen}\thanks{shenrf3@mail.sysu.edu.cn}
\affiliation{School of Physics and Astronomy, Sun Yat-Sen University, Zhuhai, 519082, China}
\author{Christopher D. Matzner}
\affiliation{Department of Astronomy and Astrophysics, University of Toronto, Toronto, ON M5S 3H4, Canada}
\author{Andrew W. Howard}
\affiliation{Department of Astronomy, Caltech, Pasadena, CA 91125, USA}
\author{Wei Zhang}
\affiliation{Key Laboratory of Optical Astronomy, National Astronomical Observatories, CAS, Beijing 100101, China}

%%%%%%%%%%%%%%%%%%%%%%%%%%%%%%%%%%%%%%%%%%%%%%%%%%%%%%%%%%%%%%%%%%%%
\begin{abstract}

At about 70 solar masses, the recently-discovered dark object orbited by a B-type star in the system LB-1 is difficult to understand as the end point of standard stellar evolution, except as a binary black hole (BBH). LB-1 shows a strong, broad H-alpha emission line that is best attributed to a gaseous disk surrounding the dark mass. We use the observed H-alpha line shape, particularly its wing extension, to constrain the inner radius of the disk and thereby the separation of a putative BBH. The hypothesis of a current BBH is effectively ruled out on the grounds that its merger time must be a small fraction of the current age of the B star.  The hypothesis of a previous BBH that merged to create the current dark mass is also effectively ruled out by the low orbital eccentricity, due to the  combination of mass loss and kick resulted from gravitational wave emission in any past merger. We conclude that the current dark mass is a single black hole produced by the highly mass-conserving, monolithic collapse of a massive star.  

\end{abstract} 

%%%%%%%%%%%%%%%%%%%%%%%%%%%%%%%%%%%%%%
\section{Introduction}

\citet{liu19} recently discovered the LB-1 system during a radial velocity (RV) monitoring search for spectroscopic binaries containing black hole (BH) candidates.  Located toward the Galactic anti-center, LB-1 consists of a B-type star with a RV period of $P=79$ d. Spectroscopic analysis suggests that its luminous component is a sub-giant with a mass $M_B \approx 8 M_{\odot}$, a radius of $9 R_{\odot}$, a metallicity of $1.2 Z_{\odot}$, and an age $t_B \approx$ 35 Myr. The fit to the spectral energy distribution from $U$, $B$ and $V$ photometry yields a distance of 4.2 kpc. The fit to its RV curve gives a semi-amplitude $K_B = 52.8 \pm 0.7$ km/s and an eccentricity $e = 0.03 \pm 0.01$, thus a mass function $PK_B^3/2\pi G=  
1.20 \pm 0.05 M_{\odot}$, suggesting the unseen companion has a mass of at least $ 6 M_{\odot}$ (for edge-on view), making it a BH candidate. 

\begin{figure}[h]
\begin{center}
\includegraphics[width=9cm, angle=0]{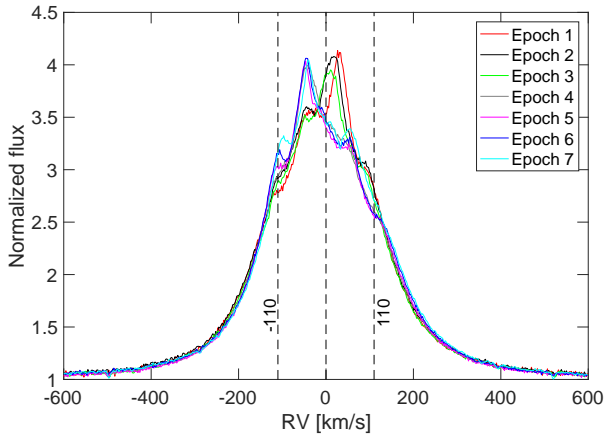}
\caption{Keck high-resolution multi-epoch data of the H$\alpha$ emission line shape from LB-1. The data set are used in \cite{liu19}}.    	\label{fig:keck-7}
\end{center}
\end{figure}

More importantly, LB-1 shows a prominent, broad (FWHM= 240 km/s) H$\alpha$ emission line (shown in Figure \ref{fig:keck-7}) which moves in anti-phase with the B star at a much smaller semi-amplitude $K_{\alpha}= 6.4 \pm 0.8$ km/s. These properties rule out a circum-binary nebula or a disk associated with the B star as the origin of the H$\alpha$ line, but they are consistent with a gaseous disk around the BH candidate \citep{liu19}. In this case, RV amplitude ratio $K_B/K_{\alpha}= M_{BH} / M_B$ immediately gives a BH mass $M_{BH}= 68^{+11}_{-13} M_{\odot}$, requiring a near-polar inclination $i \approx 15^{\circ}$-$18^{\circ}$ \citep{liu19}.

Such a large BH mass poses a challenge to current theories about BH formation and final-stage evolution of massive stars, as \cite{leung19} find that the pulsational pair-instability mass ejection before the final supernova explosion sets a maximum mass of BHs to be $\sim 50$ $M_{\odot}$.

One attractive possibility that could partially alleviate these challenges is that the primary in LB-1 is not a single, but a binary of BHs (a BBH). This would have interesting implications for gravitational wave sources. In this \textit{Letter} we consider this possibility and put constraints on it from the observed properties of LB-1's H$\alpha$ emission line (\S \ref{sec:line}), as well as from the consequence of any prior BH merger event (\S \ref{sec:merger}). Conclusions are given in \S \ref{sec:conclude}.

%%%%%%%%%%%%%%%%%%%%%%%%%%%%%
%%%%%%%%%%%%%%%%%%%%%%%%%%%

%%%%%%%%%%%%%%%%%%%%%%%%%%%%%%%%
\section{Constraint from the H$\alpha$ line}	\label{sec:line}

The possibility of a current BBH is strongly constrained by the maximum coalescence time implied by the inner radius of H$\alpha$ disk, which in turn is constrained by the shape of the H$\alpha$ line. We begin with a general consideration of the kinematic information conveyed by a generic disk emission line, before constraining the inner and outer radii of the disk in LB-1. 

%%%%%%%%%%%%%%%%%%%%%%
\begin{figure}[h]
\begin{center}
\includegraphics[width=7.5cm, angle=0]{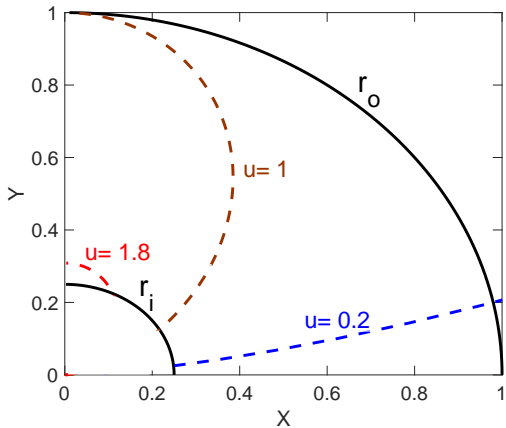}
\caption{The illustration showing one quarter of the H$\alpha$-emitting disk. The colored dashed lines mark the regions with constant line-of-sight velocity component $u$. The disk outer radius $r_o$ is set to be 1, and $u$ is set to be 1 for $\theta= \pi/2$ at $r_o$.}    \label{fig:domain}
\end{center}
\end{figure}
%%%%%%%%%%%%%%%%%%%%%%%%%%%

%%%%%%%%%%%%%%%%%%%%%%%%%%%%%%%%%%%%%%%
\subsection{Kinematically broadened line shape}

We assume the emission line's thermal width is much smaller than the kinematically broadened line width, so the intrinsic line shape can be taken as a $\delta$-function. Consider a thin Keplerian disk with an inner radius $R_i$, an outer radius $R_o$, and an inclination angle $i$. On the disk's outer rim the largest line-of-sight velocity component (radial velocity) is $v_0= (GM/R_o)^{1/2} \sin i$, where $M$ is the central object's mass. Let $\theta$ be the azimuthal angle on the disk plane, and $\theta=0$ corresponds to the projected location of the observer's line of sight. 

Let $r=R/R_o$ be the dimensionless radius, and $u(r,\theta)$ be the local radial velocity on the disk, normalized by $v_0$; then
\beq		\label{eq:u}
u= r^{-1/2} \sin \theta.
\eeq
Note that %the range of $u$ is 
$0 \leq u \leq u_{\rm max}$ where $u_{\rm max} = r_i^{-1/2}$.

The observed line emission flux $F$ within the velocity interval $(u, u+du)$ is \citep{smak69,smak81,huang72,horne86}
\beq    \label{eq:F(u)}
F(u) du = \iint\limits_{D(u)} j(r)\, r dr d\theta,
\eeq
where $j(r)$ [erg s$^{-1}$ cm$^{-2}$] is the line emissivity function; we will simply take it as a power law function $j(r) \propto r^{-a}$. The integral region $D(u)$ on the disk is given by 
\beq
u \leq r^{-1/2} \sin \theta \leq u + du.
\eeq 
Note that $F(u)$ is symmetric at $\pm u$. Also, the two quarters $0 < \theta < \pi/2$ and $\pi/2 < \theta < \pi$ contribute equally to $F(u)$, so we consider the first quarter only in computing the integral, then later multiply the result by 2. The dashed line in Figure \ref{fig:domain} illustrate $D(u)$ for three $u$ values, respectively. 

To carry out the double integral, we first do the integral over $\theta$ with $r$ being fixed.  Change the differential variable using eq. (\ref{eq:u}): 
\beq
d\theta= \frac{d(\sin \theta)}{\sqrt{1-u^2 r}} =\frac{r^{1/2}\,du}{\sqrt{1-u^2 r}} ,
\eeq
then eq. (\ref{eq:F(u)}) becomes
\beq
\begin{split}
F(u) du &= 2 \int j(r) dr \int_{u r^{1/2}}^{r^{1/2}(u+du)} \frac{d\sin \theta}{\sqrt{1-u^2 r}} \\
&= 2\,du \int \frac{j(r) r^{3/2}\,dr}{\sqrt{1-u^2 r}}.
\end{split}
\eeq
Thus,
\beq
F(u)= 2 j(r_o) \int_{r_i}^{\min(1/u^2, 1)} \frac{r^{3/2-a}\,dr}{\sqrt{1-u^2 r}}.
\eeq
Note that when $u > u_{\rm max}$, the upper limit of the integral will be $< r_i$, then one has to set $F(u)=0$.  For ease of computation, change the variable: let $x \equiv r^{1/2} u$. Then,
\beq		\label{eq:Fu}
F(u)= 4 j(r_o) u^{2a-5} \int_{x_i}^{x_o} \frac{x^{4-2a}\, dx}{\sqrt{1-x^2}},
\eeq
where $x_i= r_i^{1/2} u$ and $x_o= \min(1, u)$. Again, when $u \geq u_{\max}$, $F(u)$ shall be set to 0.

%%%%%%%%%%%%%%%%%%%%%%%%%%%%%
\begin{figure}[h]
\begin{center}
\includegraphics[width=8.5cm, angle=0]{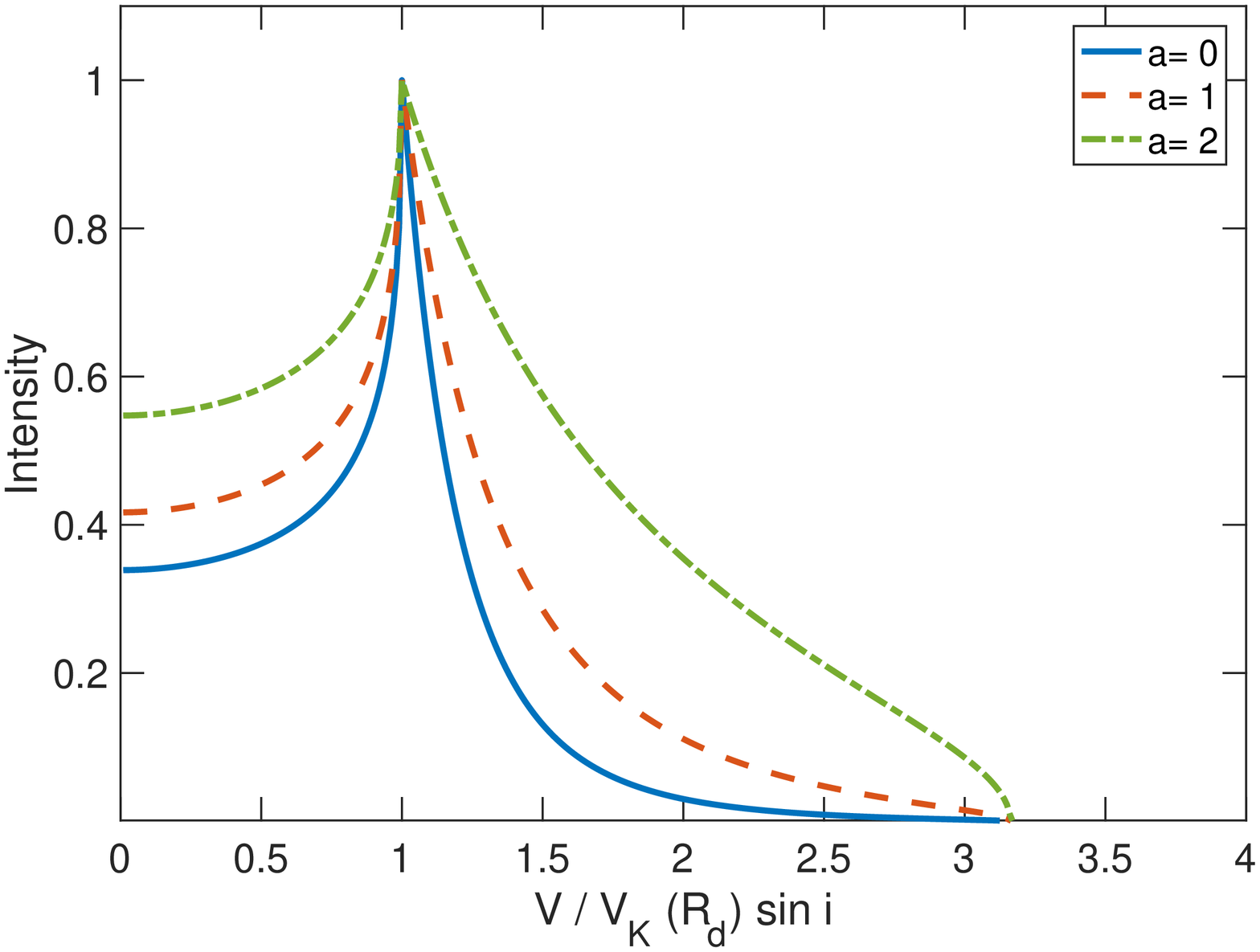}
\includegraphics[width=8.5cm, angle=0]{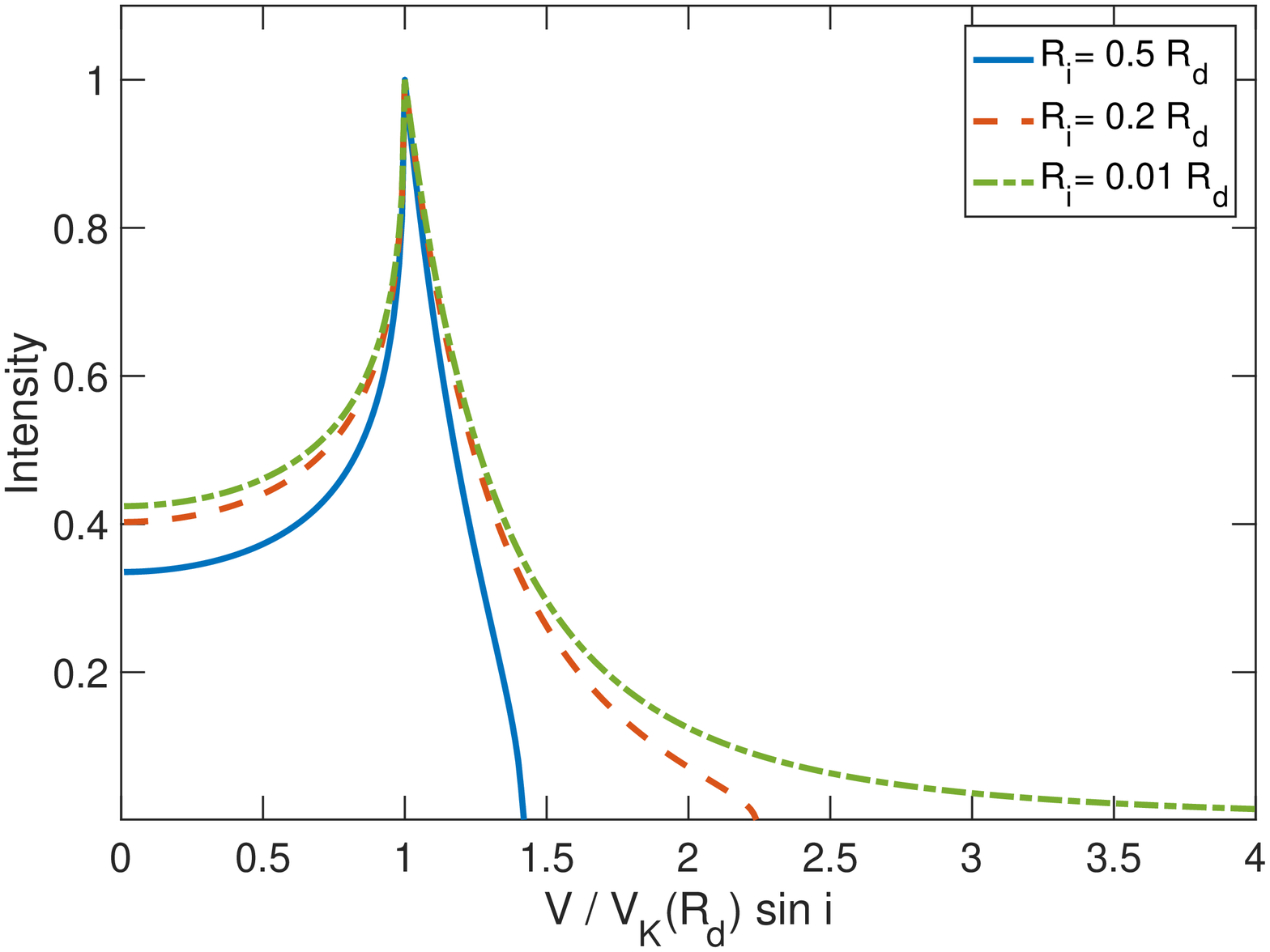}
\caption{The dependence of the line shape, computed from equation (\ref{eq:Fu}), on the emissivity's radial index $a$ (\textit{top}) and on the disk's inner radius $r_i$ (\textit{bottom}).}    \label{fig:depend}
\end{center}
\end{figure}
%%%%%%%%%%%%%%%%%%%%%%%%%%%

%%%%%%%%%%%%%%%%%%%%%%%%%%%%%%%%%%%%%%%%%%
\subsection{Model parameter dependence}

Equation (\ref{eq:Fu}) is for the case in which the disk is optically thin to the line photons. It produces a well known double-peak line profile which peaks symmetrically at $u= \pm 1$ (i.e., RV$= \pm v_0$, e.g., \citealt{huang72}). The shape of the line wings is determined by how the emissivity varies radially and how far the disk extends inward. \cite{smak69,smak81} investigated the line profile's dependence on $a$ and $r_i$, finding that $a$ controls the slope of the wings and $r_i$ determines their extension (as $u_{\rm max}= r_i^{-1/2}$), as we reproduced in Figure \ref{fig:depend}.

LB-1's emission line shows a central peak with shoulders, the so-called ``wine-bottle'' shape seen in some Be stars. According to \cite{hummel92} and \cite{hummel94}, this type of line shape is due to a combination of non-coherent scattering and kinematic broadening, and appears only at low inclination angles ($i \sim 5^{\circ} - 20^{\circ}$) and only when the disk is optically thick at the line center. For optically thin cases, the wine-bottle shape is replaced by the classical double-peak profile. 

Reproducing the wine-bottle shape seen in LB-1 requires a three-dimensional radiative line-transfer computation, which is beyond the scope of this Letter. Nevertheless, results from such a computation by \citeauthor{hummel94} (\citeyear{hummel94}, their Fig. 7) show that, for disks with the same geometrical properties, the RV positions of the shoulders of the wine-bottle shape for the optically thick case are the same as those of the double peaks in the optically thin case. Therefore, the shoulder positions still indicate the outer disk radius. Furthermore, the physical correspondence of the wing extension and slope with $r_i$ and $a$, respectively, is likewise preserved. 

%%%%%%%%%%%%%%%%%%%%%%%%%%%%%
\begin{figure}[t]
\begin{center}
\includegraphics[width=8.5cm, angle=0]{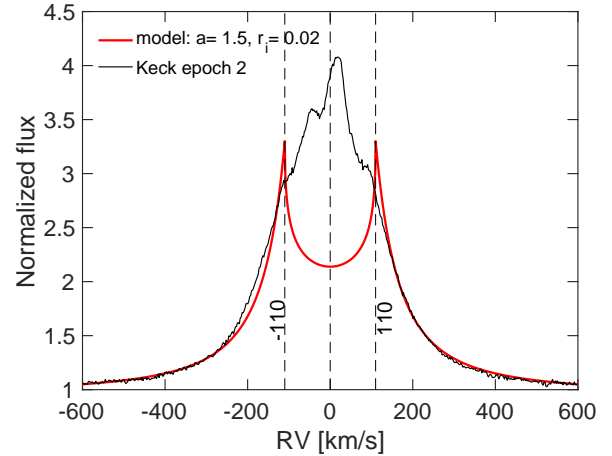}
\caption{Model fit of equation (\ref{eq:Fu}) to the Keck epoch 2 data.}    \label{fig:fit}
\end{center}
\end{figure}
%%%%%%%%%%%%%%%%%%%%%%%%%%%%%

%%%%%%%%%%%%%%%%%%%%%%%%%%%%%%%%%%%%%
\subsection{Constraint on disk inner radius}

In Figure \ref{fig:fit} we show a model fit to LB-1's line shoulder positions and the wing shape. We used the data from one observation epoch only, because the shoulder position and wind shape do not vary much among epochs (see Figure \ref{fig:keck-7}). The observed RV is $v= v_K \sin i$, where $v_K= \sqrt{GM/R}$. In modeling we have taken the dark primary mass $M= 68 M_{\odot}$ and the inclination $i= 16^{\circ}$ reported in \cite{liu19}. 

For the observed RVs of line shoulder $v_{\rm sho}= 110$ km s$^{-1}$ and line extension $v_{\rm ext} \ge 600$ km s$^{-1}$, the outer and inner radii of the H$\alpha$-emitting region of the disk are
\beq
R_o= 0.38\, \mbox{AU},~~~ R_i \le 0.013\,\mbox{AU} = 2.8\,R_{\odot},
\eeq 
respectively. Note that 0.013 AU is only an upper bound for the disk inner radius $R_i$, as the real $R_i$ could be even smaller if the innermost region is ionized. 

The inferred inner disk radius strongly constrains the possibility that the central dark primary is composed of a binary of BHs with a total mass $M$. Due to gravitational interaction, such a binary would truncate the circum-binary disk at a radius that is 1.7 times the semi-major axis $a_b$ of the binary when the binary orbit is circular; a higher eccentricity would cause a larger inner gap to the disk \citep{arty94}. Therefore, $a_b \approx R_i /1.7 \le  1.6 \, R_{\odot}$.  Assuming a circular orbit, the in-spiral time of such a binary due to gravitational wave emission is \citep{peters64}
\beq
\begin{split}
T(a_b) & =  \frac{5 c^5 a_b^4}{256 G^3 M^2 \mu} \lesssim 1.3\times10^4 \\
&  \times \left(\frac{a_b}{1.6 \, R_{\odot}}\right)^4 \left(\frac{68\,M_{\odot}}{M}\right)^3 \left(\frac{M/4}{\mu}\right) \mbox{yr},
\end{split}
\eeq 
where the reduced mass $\mu$ takes a value of $M/4$ for the equal-mass case of the binary. A higher eccentricity would make this time even shorter. 

This in-spiral time is characteristically more than 3 orders of magnitude shorter than the B star age $t_B \sim 35$ Myr, which strongly disfavors the possibility of a current central black hole binary.  Indeed, the system would be obscured by its proto-stellar envelope if it were younger than $\sim 10^5$ years.   A non-equal binary does not alleviate this constraint, as the mass ratio would have to exceed $10^3$ to make the in-spiral time comparable to the lifespan of a B star. 

%%%%%%%%%%%%%%%%%%%%%%%%%%%%%%%%%%%%%
\section{A past merger of binary black holes}		\label{sec:merger}

Could it be that the merger of a binary of BHs (BBH) had happened long ago? The  low observed eccentricity of the LB-1 system severely limits this possibility, as the effects of gravitational radiation  would tend to excite much higher eccentricities, or unbind the system entirely. 

\begin{figure}
\begin{center}
\includegraphics[width=7.5cm, angle=0]{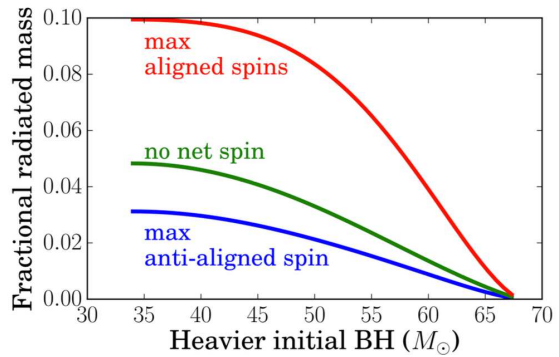}
\caption{The fractional radiated mass during a BBH merger as a function of the primary BH mass, using the model of \citet{barausse12} and assuming a total mass of $68\,M_\odot$. The red, green, and blue lines represent maximal spins aligned with the BBH orbital axis, no net spin, and maximally anti-aligned spins, respectively.}    \label{fig:Erad}
\end{center}
\end{figure}

First, consider the loss of total mass in a BBH merger due to radiated energy. \citet{liu19} infer that  the gravitating mass of LB-1 cannot have changed by more than 4\%  in a single orbit of the B star, without increasing the eccentricity above its observed value.  \citet{barausse12} present a highly accurate formula for the radiated energy, which we display in Figure \ref{fig:Erad} for the cases of maximal spins aligned or anti-aligned with the BBH orbit, as well for the case of no net spin alignment, as functions of the greater component mass  of the BBH.  From this, it is clear that, for less than 4\% of the mass to be radiated and both components to be less than $50\,M_\odot$, it requires some spin-orbit anti-alignment ($\tilde{a}<0$ in Barausse's terminology), although there is a narrow window around 47 $M_\odot$ in which no spin is necessary. 

%%%%%%%%%%%%%%%%%%
%%%%%%%%%%%%%%%%%%%%%%%

Note, however, that gravitational wave recoil is a generic outcome of BBH mergers, which tends to introduce a high eccentricity to system, or might even dissociate it. For a circularly orbiting binary of Schwarzschild BHs, \cite{fitchett83} provides a quasi-Newton estimate for the recoil velocity of the merged BH, which is equivalent to
\beq
v \approx  1000 \frac{q^2(1-q)}{1+q} ~\mbox{km/s}, 
\eeq 
which takes a maximum of about 91\,km/s when the mass ratio $q=(\sqrt{5}-1)/2$. 
Advanced calculations including numerical relativity simulations taking into account black hole spins suggest the recoil velocity is typically $v \approx 10^{-3} c= 300$\,km/s \citep{gerosa18}; and in the extreme case of maximal anti-aligned spins in a binary with $q=0.623$, the recoil can reach 526\,km/s \citep{healey14}. Considering that the circular velocity of the LB-1 system is about 100\,km/s \citep{liu19},  a kick of only a few km/s would create a greater eccentricity than what is observed.  Although it is possible to arrange for zero kick, for instance by tuning the initial BBH so that it is point-symmetric, many of these arrangements are ruled out by the radiated mass constraint (see above and Figure \ref{fig:Erad}). 

%%%%%%%%%%%%%%%%%%%%%%%%%%%%%%%
\section{Conclusions}		\label{sec:conclude}

We conclude that it is extremely unlikely that the LB-1 dark object is currently a black hole binary, or is the remnant of a merger of black holes. The only alternative is that it is a single black hole, formed by direct collapse in an event that shed very little mass.  Its existence is therefore a clear challenge to the prediction of envelope ejection caused by pair-instability pulsations, as was already stated in \cite{liu19}. The findings presented here make it an even stronger case. The single-star collapse scenario for LB-1 might be possible only when a reduction of the stellar wind loss is enforced, as was shown in \cite{belczynski19}, albeit with other difficulties there.  

We note, moreover, that no significant \textit{sudden} mass loss could have occurred since the system's orbit is currently circular, presumably having experienced tidal circularization, e.g., during a giant phase of the primary star (as inferred in Wolf-Rayet/O-star binaries: \citealt{monnier99}).  This includes any pulsation-driven mass loss events, as well as the stellar collapse itself.  

The latter is especially interesting in light of the fact that circularization would have spun up the giant's envelope prior to its collapse, leading to the creation of a highly super-Eddington accretion disk during the collapse around the nascent BH.  Such disks are thought to re-eject significant fractions of the accreting matter in the form of a disk wind, which may expel additional envelope material \citep{feng10}.  Clearly in order to maintain a very low orbital eccentricity, none of these effects removed any significant fraction of the object's mass in this case. 

\acknowledgments
RFS thanks the hospitality of Canadian Institute of Theoretical Astrophysics where part of this work was carried out. 

%%%%%%%%%%%%%%%%%%%%%%%\begin{references}%%%%%%%%%%%%%%%%%%%%%%

%%%%%%%%%%%%%%%%%%%%%%%%%%%%%%%

\end{document}